\documentstyle{aa}
\input epsf
\begin{document}
\thesaurus{06           
           (03.20.2;            
            08.02.1;            
            08.09.2 HD144217;   
            08.16.2;            
            08.12.2)}           
\title{Preliminary results of dark-speckle stellar coronography}
\author{A. Boccaletti\inst{1} \and A. Labeyrie\inst{2} \and R. 
Ragazzoni\inst{3}}
\offprints{A. Boccaletti}
\institute{DESPA, Observatoire de Meudon, 5 place Jules Janssen, 92195 Meudon, France, Boccalet@despa.obspm.fr \and Coll{\`e}ge de France \& Observatoire de Haute Provence 04870 Saint Michel l'Observatoire, France,
Labeyrie@.obs-hp.fr \and Astronomical Observatory of Padova, vicolo 
del'Observatorio 5, I-35122 Padova, Italy, Ragazzoni@astrpd.pd.astro.it
}
\date{Received 2 October 1997 ; accepted 22 January 1998}
\maketitle
\begin{abstract}
The dark-speckle method (\cite{ref9}) combines features of speckle
interferometry and adaptive optics to provide images of faint circumstellar 
material.  We present preliminary results of observations, and simulations 
concluding to the feasibility of exo-planet imaging from the ground.  
Laboratory simulations with an avalanche photodiode indicate the detectability 
of a stellar companion of relative intensity  $10^{-6}$ at 5  Airy radii from 
the star. New, more general, expressions for the signal-to-noise ratio and 
integration time are given. Comparisons with direct long-exposure imaging 
indicate that the method improves the detectability of circumstellar 
nebulosity, faint companions and planets.
\end{abstract}
\section{Introduction}
Extra-solar planets (Mayor \&Queloz 1995, Marcy \& Butler 1996) 
can, in principle, be seen in ground-based images using the
light cancellation in dark speckles to remove the halo of star light (Labeyrie 1995).
Long-exposures with high-performance adaptive optics and a 
correction of the seeing-induced shadow pattern on the telescope's pupil are 
also proposed by Angel  (1994). A less extreme atmospheric correction is needed
 with the dark-speckle method.  Both methods are 
expected to reach the $10^{-9}$ relative intensity needed to detect a  
Jupiter-like planet near its parent star. Related space-based techniques
 are also considered  (\cite{ref12}, \cite{ref5}). In the longer term, 
resolved images of detected exo-planets will in principle be obtainable even 
from the ground with long-baseline interferometric arrays (\cite{ref10}). \\
The "dark-speckle" method exploits the light cancellation effect occuring in 
random coherent fields according to the Bose-Einstein statistics. Although 
adaptive optics  can reconstruct the Airy peak, and possibly the first few 
rings, in the focal image of a bright star, the degree of "seeing" correction
 which it provides cannot be good enough to remove the fluctuating speckles in
 the surrounding zone of the Airy pattern.  Coronagraphic devices 
(\cite{ref11}, \cite{ref4}, \cite{ref15}, \cite{ref13}, Gay \& Rabbia 1996, 
Roddier \& Roddier 1997)
can remove the steady and organized part of the 
straylight, i.e. the first few Airy rings if they emerge from the  boiling 
speckled halo, and even if they are burried but remain detectable with the 
dark speckle analysis.\\ 
One is therefore left with  the problem of extracting as much
information as possible from this  speckled halo.  Destructive interference 
in the star light occasionally causes a dark speckle to appear here and there.
  When this happens at the position of the planet's own faint Airy peak, the
 darkening cannot be as deep as elsewhere. The planet's  Airy peak, also 
restored by the adaptive optics, indeed has a rather stable intensity, which 
adds to the star's local intensity. The intensity histogram is 
therefore locally distorted, and suitable algorithms can display the local
distorsions in the form of a cleaned image. 
The exposures must be shorter than the turbulence life time. With a large 
telescope and hours of integration, a planet, typically $10^9$ times fainter 
than the parent star, is expected to become visible.
\section{Signal to noise ratio}
To calculate the method's sensitivity in a more general way than done by Labeyrie (1995), we first follow his derivation to the point where he calculates the signal-to-noise ratio.\\
The different photon distribution from the star and the planet defines the signal-to-noise ratio $SNR$, and according to the central limit theorem:

\begin{equation}
nP_*(0)[1-P_0(0)]=SNR\sqrt{nP_*(0)}
\end{equation}
where $P_*(k_*)$ and $P_0(k_0)$ are the probabilities to detect $k_*$ and
$k_0$ photons originating respectively from the star and 
from the planet, per pixel in a short exposure, and where $n$ is the total number of short exposures. Replacing the values of $P_*(0)$ and $P_0(0)$ in Eq. (1), we obtain:
\begin{equation}
SNR=(1-e^{-\overline{k_0}})\sqrt{n\over 1+\overline{k_*}}
\end{equation}
The calculation of signal to noise ratio given in Labeyrie 1995 assumed
$\overline{k_*}\gg 1$, which is unnecessarily restrictive, and
unrealistic in some of the cases of interest. As suggested by one of us
(RR), a more general analysis can be made under the assumption
 that $\overline{k_0}\ll 1$ and $\overline{k_*}\gg\overline{k_0}$. Equation (2)  then becomes:
\begin{equation}
SNR\approx \overline{k_0}\sqrt{n\over 1+\overline{k_*}}
\end{equation}
if $j$ is the number of pixels per speckles, thus:
\begin{equation}
(SNR)^2\approx {{n\over j}(j\overline{k_0})^2\over
j+j\overline{k_*}}={n'\overline{K_0}^2\over j+\overline{K_*}}
\end{equation}
Where $\overline {K_0}$ and $\overline {K_*}$ are the number of photons per
speckle in a short exposure, respectively for the planet and the star.\\
The variables used in \cite{ref9} were:\\
$n'=${\Large${T\over t}$}, {\Large$\;\;{\overline {K_*} \over
\overline{K_0}}$}$=${\Large${R\over G}\;\;$} and $\;\;G\overline {K_*}=tN_*$\\
where $T$ is the total integration time, $t$ the short exposure time,
$R$ the star/planet brightness ratio, $G$ the gain of the adaptive optics
or the ratio between the Airy peak and the halo of speckles, $N_*$ the total number of photons per second detected from the star.\\
It provides a new expression for the $SNR$:
\begin{eqnarray}
SNR=\overline {K_0}\sqrt{n'\over j+\overline {K_*}} & = & {tN_*\over R}\sqrt
{n'\over j+{tN*\over G}}\nonumber \\
 & = & {N_*\over R}\sqrt{tT\over j+{tN_*\over G}}
\end{eqnarray}
The sampling $j$ should be fine enough to exploit the darkest parts of the 
dark speckles, for a given threshold of detection $\epsilon$, linking the 
performance of the adaptive optics ($G$) and the brightness ratio ($R$).
The intensity across a dark speckle may be coarsely modelled as a cosine 
function of the position $\rho$ by:
\begin{equation}
I(\rho)=I_h\left(1-\cos{2\pi\rho\over \sqrt j}\right)
\end{equation}
where $I_h$ is the mean intensity and $\sqrt j$ the speckle size.
The intersection between $I(\rho)$ and the line $y=\epsilon I_h$ gives $s$, the
size of the pixel over which the light is integrated. It limits the minimal measurable intensity $\epsilon I_h$ (with $0<\epsilon \ll 1$).
A detailed calculation gives $s\approx 1.27\sqrt {\epsilon}${\Large${\lambda\over
D}$}.\\
We can assume that $\epsilon=G/R$.
Indeed, if $I_0$ is the intensity of the Airy peak,
$\epsilon=${\Large${\epsilon
I_h\over I_h}$}$=${\Large${\epsilon I_h\over I_0}{I_0\over
I_h}$}$=${\Large${G\over R}$}\\
We also assume in the following that a planet can be seen if its own intensity is higher than $\epsilon I_h$.\\
Now we are able to calculate a value of $j$:
\begin{equation}
j={\left(\lambda /D\right)^2\over s^2}={\left(\lambda /D\right)^2\over
{1.27^2\epsilon \left(\lambda /D\right)^2}}=0.62{R\over G}
\end{equation}
\begin{figure}[t]
\centerline{
\epsfbox{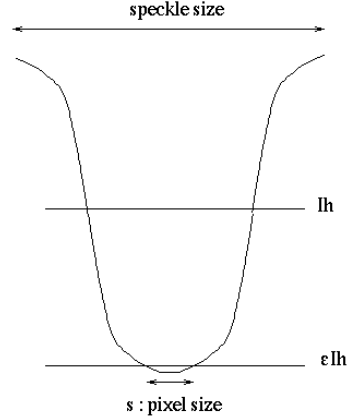}}
\caption[]{Shape of a dark speckle according to Eq. (6).
If $I_h$ is the mean intensity, integration on a pixel of size $s$ yields an
intensity $\epsilon I_h$.}
\end{figure}
Recent numerical simulations (\cite{ref3}) have shown that $R/G$ should not exceed $10^3$ to retain a reasonable value of the sampling parameter $j$.\\
The recorded level is $I_h$, but with dark speckles this residual level 
decreases to $\epsilon I_h$, where $\epsilon$ depends mainly on the adaptive 
optics performance.\\
With the value of $j$ obtained, the $SNR$ expression becomes:
\begin{equation}
SNR={N_*\over R}\sqrt{tT\over 0.62{R\over G}+{tN_*\over G}}
\end{equation}
Solving for $R$, leads to a third-degree equation:
\begin{equation}
0.62R^3+tN_*R^2={GtTN_*^2\over (SNR)^2}
\end{equation}   
The Cardan method gives a single positive solution from which we derive a 
final expression for the integration time.
\begin{equation}
T=\left({SNR\over N_*}\right)^2{R\over Gt}(0.62R^2+tN_*R)
\end{equation}
With the following values : $R=10^9$; $D=8m$; $G=10^6$; $SNR=5$; $m_v=2.5$;
$q=0.2$; $\Delta\lambda=100nm$ and $t=20ms$, we find $2.7$ hours of
integration, i.e. $50\%$ more than the result given by Labeyrie.
\\
Angel's discussion (1994) of the long-exposure method leads him to the following 
expression:
\begin{equation}
R={G\over SNR}\sqrt{T\over \Delta t_{opt}}
\end{equation}
where $\Delta t_{opt}$ is the optimum short exposure time. Using
again the same values, Eq.(11) gives a limiting brightness ratio of
$1.4.10^8$ and the typical brightness ratio of $10^9$ would be reached in $140$ hours. The long-exposure approach would in principle be more sensitive if
the adaptive optics and shadow pattern compensation could be made extremely
good.  It is however less sensitive with current levels of adaptive performance.\\
In fact, the difference between the dark-speckle and direct-long-exposure
methods is more 
subtle, and both work in different regimes. A critical value of the photon 
stellar flux ($N_{*c}$) can be easily calculated by equalizing Eq.(10) and 
Eq.(11) for the same value of the total integration time ($T$). It leads to a
 second degree equation in $N_*$, with a single positive root:\\
\begin{equation}
N_{*c}={G+\sqrt{G^2+4\times 0.62GR}\over 2t}
\end{equation}
Therefore, if the photon flux is above $N_{*c}$, the dark-speckle method is 
more efficient than the long exposure and, below this limit the direct imaging 
is better. $N_{*c}$ strongly depends on the adaptive optics. Let us take
 a numerical exemple. If the goal is to reach a $10^9$ brightness ratio with a 
gain of $10^6$ and $20ms$ exposures, the photon flux must be higher than 
$1.27.10^9 photons/s$, which can be achieved with a large aperture and wide 
bandwidth ($9.3.10^9$ph/s for $m_v=0$, $D=2.4m$, $\Delta\lambda=100nm$ and 20\% efficiency). However, for a space telescope with adaptive optics, the speckle 
lifetime is under control with values of the order of $1s$ for exemple. 
Thus, $N_{*c}$ is decreased to $2.54.10^7 photons/s$, which is less restrictive.
However, the telescope should not be so large as to provide a partially
resolved image of the star, since it would fill-in the dark speckles.
\section {Dark speckle lifetime}
Because the number of photon-events detected per pixel in each exposure is critical, these exposures should be as long as possible without degrading too much the darkness of the dark speckles. The optimal value is obviously shorter than the usual speckle lifetime considered by Roddier, Gilli \& Lund (1982). 
A tentative lower limit can be estimated by linearly scaling the speckle lifetime in proportion to the dark speckle size defined by Eq.(7). It leads to impossibly short exposures for large $R$ values.
However, since the adaptive optics decreases the wave disturbance, it increases
 markedly the speckle lifetime at positions close to the Airy peak,  
depending upon the factor $n{\lambda\over D}$, where $n$ is the number of 
actuators across the pupil containing $n^2$ of them (\cite{ref20}). 
The optimal exposure time is therefore dependant upon the adaptive optics performance.
\section{Simulation and results}
\begin{table}
\caption[]{Brightness ratio and $SNR$ obtained with a photon-counting avalanche
photodiode, for two values of the sampling 
parameter $j$. The number of zero-photon events was counted on 250000 short 
exposures of $100\mu s$, totalling $25s$ of integration. The $SNR$ was 
calculated from Eq.(1).}
\begin{center}
\begin{tabular}{|c|c|c|c|c|c|}\hline
$j$ & $R_{max}$ & $R$ & 0 ph. & 0 ph. (star & $SNR$ \\
 & & & (star) & +planet) & \\ \hline
 & & 15000 & 136796 & 110377 & 71.4 \\ \cline{3-6}
80 & 440000 & 150000 & 87672 & 85546 & 7.2 \\ \cline{3-6}
 & & 360000 & 82340 & 80625 & 5.9 \\ \hline
144 & 790000 & 560000 & 111044 & 107946 & 9.3 \\ \cline{3-6}
 & & 950000 & 155959 & 150017 & 4.0 \\ \hline
\end{tabular}
\end{center}
\end{table}
To assess the dark-speckle method we did a laboratory simulation using a
single-pixel photon-counting detector, in the form of an avalanche
photodiode.
The star and planet were simulated by two He-Ne lasers, with
adjustable attenuators. 
A Lyot-type coronagraph permitted to remove the star's
Airy peak and rings, thus decreasing the local halo intensity 10
to 15 times. Artificial ``seeing'' was generated with a moving
scatterer, selected to provide a Strehl ratio approaching that typical of
current adaptive optical systems.  The equivalent peak/halo gain was
$G=3.4.10^3$. The flux of the central star was $44.10^6$
photons/s. Calculating an histogram of  the detected photon events, we
determined the
$SNR$ by comparing the number of zero-photon events with the planet
turned on and off.
As listed in Table 1, the results strongly depend
on the sampling parameter $j$. They are consistent with Eq.(7) which gives
the maximum brightness ratio ($R_{max}$) theoretically reachable. 
In these laboratory tests, the dark-speckle analysis outperforms the long exposure when the sampling exceeds $144$ $pixels/speckle$ $area$. 
\begin{figure}[hb]
\vspace{0.3cm}
\centerline{
\epsfbox{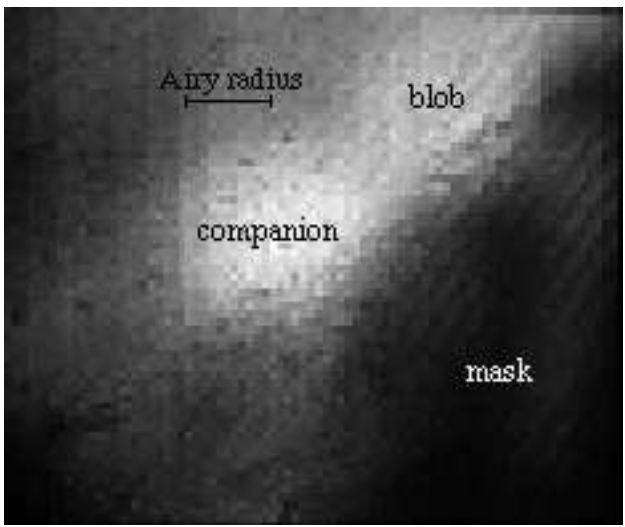}}
\caption[]{Laboratory simulation of stellar companion detection with the dark 
speckle method. The artificial companion, $966$ times fainter than the star, 
is at the center of the dark map, emerging from the halo. A coronagraphic mask, which hides the star's central peak is visible in the lower right corner. The 
blob seen in the upper right part of the field is a ``static speckle'' 
caused by permanent aberrations and removable with a reference star.}
\end{figure}
In this experiment the short exposure time ($100\mu s$) is about 100 times shorter than the speckle lifetime ($10ms$).
Available photon counting camera do not yet allow quite as short exposure.\\[0.2cm]
\begin{figure}[t]
\centerline{
\epsfbox{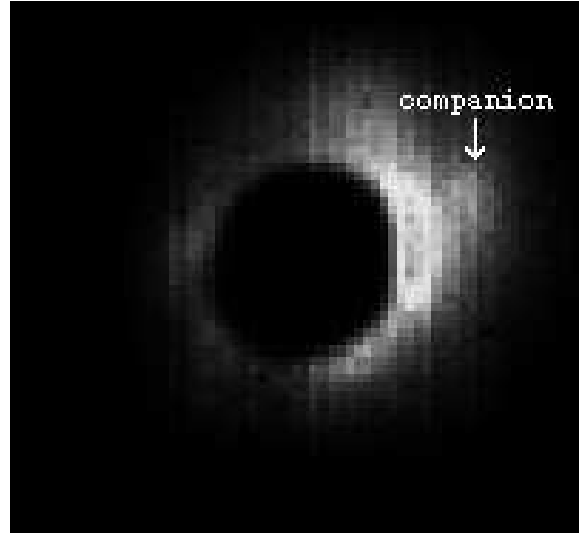}}
\caption[]{Cleaned image, generated through dark-speckle analysis, of the 
multiple star HD144217($\beta$ Sco). A companion ($\beta$ Sco B) appears
 near the masked star image. (mask diameter$=0.5''$, $F/D=1200$)}
\end{figure}
We also used the CP40
photon-counting camera developed by Foy and Blazit (\cite{ref2}). It has pixels
of $50\mu m$ and a rather low saturation level of $50000 photons/s$. Each 
pixel has a lower dark noise than an avalanche photodiode, but a much slower 
response. We used the algorithm described in Labeyrie (1995),
generating a ``dark map'' by counting, in each pixel, the number of $20ms$
exposures which contribute
 zero photon. As contributions from
successive $20ms$ exposures are accumulated in the dark map, a planet's
Airy peak is expected to emerge as a black dot among background
noise.
To obtain the cleaned image, results are displayed in positive form 
using, for example, an inverse square law. \\
The CP40 discriminates between events featuring zero photoelectron and those featuring one or more photoelectron/pixel/exposure. Adding all exposures generates  an image which brings out faint companions better than would a similarly long exposure on a CCD detector.\\
The flux of the central star was
$5.3.10^6$ photons/s, and the gain of the adaptive optics was about
$1556$. The optical system operated at
$\lambda =0.67\mu m$, $F/D=3200$, and the mask diameter was $0.34"$, i.e. it covered the central 2 rings of the Airy pattern.
The planet was located near the fifth ring of the diffraction pattern.\\
Figure 2 shows that a companion $966$ times fainter is well detected in
116 seconds. 
The $SNR$ measured on a speckle size region ($37\times 37$ pixels) is $799$, while the
dark-speckle model predict an $SNR$ of $1008$. This model does not take into
account the halo shape which can explain the $20\%$ discrepancy. 
These initial results, where the companion is brighter than the halo, are very modest with respect to the performance expected at a later stage, but have provided useful insight for improving the instrument.\\
Currently, as seen in Figs 2 and 3, the detection sensitivity is limited by the presence
of spurious blobs in the cleaned image. These are caused by static aberrations 
and coronagraphic mask effects. These residual blobs may be substracted from 
data obtained on a reference star.\\[0.2cm]
Finally, dark-speckle data have been recorded at the 152cm telescope
of  Haute-Provence, using,  during a single night, the BOA
adaptive optics system (\cite{ref19}) developped by the Office National d'Etudes
et de Recherches A{\'e}rospatiales (ONERA).
\begin{figure}[t]
\centerline{
\epsfbox{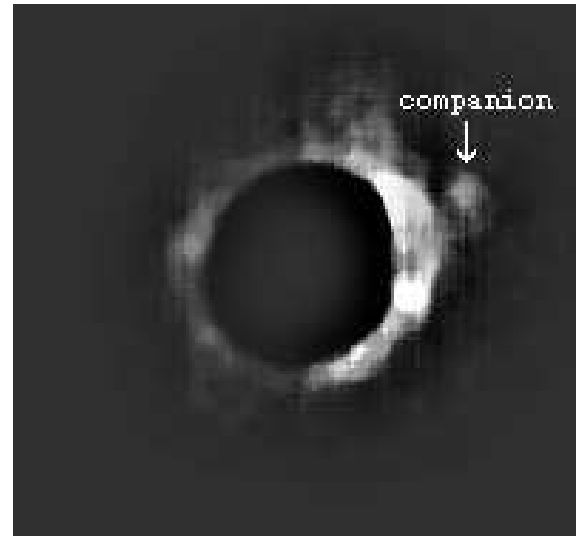}}
\caption[]{Same image enhanced with unsharp mask and high pass filter to emphasize the contrast and median filter to smooth the image at the speckle scale.}
\end{figure}
This system, optimized for visible light, reaches a Strehl ratio of 
about 0.4 at $0.6\mu m$ in long exposures and higher in short exposures. 30 
minutes of observation, with an interference 
filter centered at $0.67\mu m$ ($\Delta\lambda=100 \mbox{\AA}$), evidenced the
faint component of the spectroscopic binary HD144217 ($\alpha=16h05'26"$,
$\delta=-19^{\circ}48'18"$, $V=2.62$). On the detector, the flux from the
primary was only $11860$ photons/s.
The angular separation is about $0.45"$ with an uncertainty due to the
mask offset. Owing to the low elevation of the star, the adaptive optics gain 
was only $12$. The $SNR$ measured on $16\times 16$ pixels is $168$ and allows 
to derive
from Eq.(10) a brightness ratio of $88$, corresponding to a $4.8$ magnitude
difference. On the long exposure synthesized from the same data 
(1 photon-events analysis) the $SNR$ is very similar, but a direct imaging 
should give a $SNR$ of $42$ according to Eq.(11).
However, a recent measurement of lunar-occultation (\cite{ref6}) gives $\Delta
 m=3.3$. In this case, dark-speckle analysis should provide a very good detection
 with a $SNR$ of $756$ instead of $168$. 
Unfortunately, the Hipparcos mission failed to detect the companion, probably
because the Hipparcos satellite is unable to achieve $\Delta m>4$ with small
angular separation, which is consistent with our data.  
More observations are needed to verify the companion magnitude.\\
A continuing observing program is initiated.
\section{Conclusion}
The simulations and tentative observations lead to the following remarks:

1. One would like  to sample as densely as possible to reach the bottom of
the dark speckles, but there should be enough photons per pixel. The optimal sampling is therefore critical and we guess that it should be about 500 pixels per speckle area.

2. A fast photon-counting
camera with a low dark noise, high saturation level and many pixels is 
needed. 

3. The observations required a narrow-band filter since the diffraction and 
speckle
pattern are color-dependant. The speckles are themselves dispersed
 radially. To increase the bandwidth usable in speckle interferometry,
Wynne designed
 a chromatic lens with magnification inversely proportional to wavelength
(\cite{ref22}).
 D. Kohler built a Wynne corrector and we found it efficiently applicable to the present situation, where the speckle's
wavelength
 dependance is more nearly a linear scaling. The resulting smearing of the 
planet's peak is acceptable if the spectral band remains less than 100 nm.

4. Different types of apodisation can be achieved, using a classical Lyot
coronagraph, the interference coronagraph of  Gay \& Rabbia
(1996), or the phase-mask coronagraph of Roddier \& Roddier 
(1997). Both recent systems favor the detection of planets closer to
the central star's Airy peak. Laboratory simulations with these varied
devices are considered  to compare their respective efficiencies.
\\[0,2cm]
Our simulations verify the theoretical expressions given for the signal to noise
ratio. The $SNR$ measured from the photon-number variance (Eq.(1)), is consistent
with the $SNR$ expected from the model (Eq.(10)). In these preliminary tests,
we had to use an interference filter and low saturation level camera which
provides a weak signal. We were consequently unable to reach enough sensitivity
for detecting extrasolar-planets or even brown-dwarf companions.\\
The dark-speckle method is also applicable to space telescopes. Even
without turbulence, optical defects create  static speckles which
can be made to fluctuate with a few actuators, arranged in the form of an
active optics system, or a fast random scatterer.  We proposed a "dark-speckle camera", the Faint 
Source Coronagraphic Camera for the Hubble Space Telescope (\cite{ref5}). The 
project is reconsidered for the New Generation Space Telescope.\\
IR wavelengths are of interest for the detection of extrasolar
planets, for two reasons: the planet's contrast is improved and, 
turbulence is easier to correct at these wavelengths.     
The forthcoming developement of bidimensional sensors with low read 
noise should allow red and IR work.
\acknowledgements{We wish to thank D. Kohler and G. Knispel who made
simulations possible, as well as D. Mourard and A. Blazit for the CP40 camera
assistance. We are also grateful to the ONERA team for providing the adaptive 
optics system.}

\end{document}